\documentclass[conference]{IEEEtran}
\IEEEoverridecommandlockouts
\usepackage{cite}
\usepackage{amsmath,amssymb,amsfonts}
\usepackage{graphicx}
\usepackage{textcomp}
\usepackage{xcolor}
\def\BibTeX{{\rm B\kern-.05em{\sc i\kern-.025em b}\kern-.08em
    T\kern-.1667em\lower.7ex\hbox{E}\kern-.125emX}}

\usepackage{url}
\usepackage{tabularx}
\usepackage{booktabs}
\usepackage[ruled,vlined]{algorithm2e}
\usepackage{amsthm}
\usepackage{array}

\newcommand{\property}[2]{%
  \vspace{0.5em}
  \noindent\textbf{#1.} \textit{#2}
  \vspace{0.5em}
}

\bstctlcite{ieeectrl}

\begin{document}

\title{RaceTEE: Enabling Interoperability of Confidential Smart Contracts}

\author{
\IEEEauthorblockN{Keyu Zhang}
\IEEEauthorblockA{\textit{Department of Computer Science} \\
\textit{University of Oxford}\\
Oxford, United Kingdom\\
keyu.zhang@cs.ox.ac.uk}
\and
\IEEEauthorblockN{Andrew Martin}
\IEEEauthorblockA{\textit{Department of Computer Science} \\
\textit{University of Oxford}\\
Oxford, United Kingdom\\
andrew.martin@cs.ox.ac.uk}
}

\maketitle

\begin{abstract}
Decentralized smart contracts enable trustless collaboration but suffer from limited privacy and scalability, which hinders broader adoption. Trusted Execution Environment (TEE) based off-chain execution frameworks offer a promising solution to both issues. Although TEE-based frameworks have made significant progress, prior work has yet to fully explore contract interoperability, a critical foundation for building complex real-world decentralized applications. This paper identifies the key challenges impeding such interoperability and presents practical solutions. Based on these insights, we introduce RaceTEE, a novel framework that leverages off-chain TEE-enabled nodes to efficiently execute confidential, long-lived smart contracts with interactions of arbitrary complexity among contracts. We implement a RaceTEE prototype using Intel SGX, integrate it with Ethereum, and release it as open source. Evaluation across diverse use cases demonstrates its practicality and effectiveness.
\end{abstract}


\section{Introduction}
\label{sec:intro}
Since Ethereum emerged as a global decentralized computing platform supporting Turing-complete smart contracts with guaranteed correctness and integrity, blockchain applications have expanded significantly beyond digital currencies. However, despite the increasing adoption of smart contracts on public blockchains, several challenges remain, particularly regarding privacy and computational overhead. On-chain transparency inherently limits the handling of sensitive data, while redundant execution across multiple nodes, necessary for consensus and security, results in substantial inefficiencies.

As more real-world scenarios, including finance, healthcare, and gaming, are integrated into blockchain platforms, these limitations become more pronounced. For instance, in the blockchain-based insurance industry, companies must not only protect sensitive client information and insurance details but also perform increasingly complex calculations quickly and accurately to determine appropriate compensation.

Various techniques have been used to address these challenges, including Trusted Execution Environments (TEEs), which provide hardware-based confidentiality and integrity through isolated execution environments that are tamper-resistant and immune to host interference. By offloading smart contract execution from the large number of on-chain nodes to a limited set of TEE-enabled nodes, TEE-based off-chain execution frameworks significantly improve execution efficiency by reducing redundancy, while enabling secure and private computation with minimal overhead.

Despite their potential, current TEE-based frameworks predominantly focus on single or short-lived contracts, which limits their applicability in broader use cases. For example, Etherisc's flight-delay insurance, a well-established Ethereum-based product, often requires prolonged execution spanning thousands or even millions of blocks (spanning days or months), along with extensive interactions among multiple contracts, including seven modular components of its own as well as several external oracle and token service contracts \cite{etheriscFlightDelay}. Compressing such heavy and intricate logic into a single, short-lived contract solely for privacy preservation is highly impractical, rendering existing frameworks inadequate for complex, real-world scenarios.

Indeed, Pigaglio et al. reported that cross-contract interactions accounted for 49\% of all historical blockchain transactions, and this proportion continues to rise \cite{pigaglioExploringLocalityEthereum2023}. Furthermore, over 70\% of transactions in 2023 involved inter-contract calls, with an average of 8.94 interactions per transaction \cite{hongProphetConflictFreeSharding2023a}. Hence, effectively supporting increasingly complex and frequent inter-contract interactions among long-lived contracts has become an urgent requirement for off-chain privacy-preserving frameworks.

However, addressing privacy, efficiency, and interoperability simultaneously poses significant challenges. In addition to well-known issues identified by prior work, such as malicious host input injection, unresponsive off-chain nodes and so on, three additional critical limitations must be overcome:

\begin{enumerate}
    \item Unified Execution Environment: To support confidential contract interoperability, all contracts should run within a unified and secure execution environment rather than being executed in isolation. This environment must support authorized inter-contract function calls with data sharing, while strictly enforcing privacy boundaries to prevent unauthorized information leakage.

    \item Contract Consistent Availability: To manage complex inter-contract dependencies, the framework must ensure availability of all involved contracts, even in the presence of many unresponsive off-chain nodes. This guarantees that contracts can safely depend on one another without availability concerns.

    \item Global Transaction Order: Independent transaction execution without enforced order risks inconsistencies due to dirty reads and writes on shared contract state. The framework must ensure a globally consistent transaction execution order across off-chain nodes to maintain correctness and stability.
\end{enumerate}

To address these challenges, we propose RaceTEE. By employing dual-layer encryption with periodic key rotation, competitive execution among randomly selected nodes, and treating each block as an atomic processing unit, RaceTEE establishes a unified off-chain execution environment, provides contract-consistent availability, and ensures a consistent global transaction order. As a result, it efficiently supports complex inter-contract interactions while maintaining compatibility with existing blockchain infrastructures.

In summary, this paper makes the following contributions:
\begin{itemize}
    \item We identify key but previously unexplored challenges in enabling contract interoperability for off-chain frameworks and present practical solutions, introducing a novel design perspective for building a unified and order-deterministic off-chain execution environment.
    
    \item We propose RaceTEE, which, to the best of our knowledge, is the first TEE-based off-chain execution framework that efficiently supports long-lived and complex inter-contract interactions while ensuring high compatibility and user transparency.
    
    \item We conduct a systematic analysis of the RaceTEE protocol, demonstrating its security guarantees in terms of correctness, liveness, and privacy, even in the presence of powerful adversaries.

    \item We implement a prototype of RaceTEE, demonstrate its practicality across diverse use cases, and evaluate its performance.
\end{itemize}

The paper is structured as follows: Section~\ref{sec:work} reviews related work and examines the limitations of existing approaches in supporting contract interoperability. Section~\ref{sec:adversary} outlines the security assumptions and threat model that form the foundation of our design. Section~\ref{sec:design} presents our solutions to the identified challenges, detailing the RaceTEE architecture, workflow, and core components. Section~\ref{sec:security} provides a systematic analysis of RaceTEE’s security properties, while Section~\ref{sec:eval} describes the prototype implementation and evaluates its performance across various use cases. Finally, Section~\ref{sec:conclusion} concludes.

\section{Related Work}
\label{sec:work}
Various research efforts have explored cryptographic techniques to enhance smart contract privacy, including homomorphic encryption, multi-party computation (MPC), and zero-knowledge proofs (ZKPs) \cite{kosbaHawkBlockchainModel2016,kumaresanHowUseBitcoin2015,solomonSmartFHEPrivacyPreservingSmart2023,steffenZeeStarPrivateSmart2022,steffenZapperSmartContracts2022,zyskindEnigmaDecentralizedComputation2015}. However, these methods often encounter limitations such as significant computational overhead, restricted generality, and incomplete privacy assurances, hindering their practical deployment.

In contrast, TEE-based smart contract execution addresses these challenges by enabling confidential computation with lower overhead, enhanced scalability, and broad applicability. Some approaches, such as Secret Network and Ten (formerly Obscuro), deploy TEEs on-chain to achieve confidential smart contract execution \cite{jamesTenConfidentialSmart,thefoundationofaconfidentialdecentralizedfutureSecretNetworkGraypaper}. Nonetheless, these solutions require all blockchain nodes to be TEE-equipped and to redundantly execute contracts within TEEs, which increases overhead and expands the trusted computing base (TCB), thereby reducing efficiency and introducing additional security risks.

Several approaches adopt off-chain TEEs to support computations with short durations or limited participants. FastKitten enables private, fixed-participant multi-party computation (MPCs) rounds on Bitcoin \cite{dasFastKittenPracticalSmart2019b}, while DeCloak further enhances availability and fairness in MPCs settings \cite{renDeCloakEnableSecure2023}. PrivacyGuard focuses on policy-compliant private data sales and usage \cite{xiaoPrivacyGuardEnforcingPrivate2020}, and b-DTC facilitates secure, decentralized outsourcing of computational tasks \cite{liangBlockchainbasedPlatformDecentralized2024}—both emphasizing single-instance computations. However, these works are not designed to support generalized, long-lived contracts that are common on public blockchain platforms such as Ethereum.

Other off-chain TEE-based frameworks support private long-lived contracts but focus solely on single-contract execution without addressing inter-contract interactions. Ekiden requires developers to bind each contract to an off-chain TEE-equipped node, which uploads encrypted contract state after every transaction to support failure recovery through checkpointing \cite{chengEkidenPlatformConfidentialityPreserving2019a}. ShadowEth and TCaaS move the selection of off-chain TEE-equipped nodes on-chain, where selected nodes autonomously monitor the blockchain and pull only transactions assigned to them for execution. \cite{yuanShadowEthPrivateSmart2018,liTrustingComputingService2023}. POSE optimizes cost by relying on a pool of off-chain TEE-equipped nodes, rather than a single one, to replicate intermediate states and mitigate the risk of executor dropout \cite{frassettoPOSEPracticalOffchain2023a}. LucidTEE addresses compliance with history-based policies \cite{gaddamLucidiTEEScalablePolicyBased2023}, while EPT focuses on improving user experience \cite{duEPTEnhancingUser2023}. However, these studies predominantly focus on transactions within a single contract, without guaranteeing the availability of dependent contracts. Moreover, they execute transactions from different contracts in isolated nodes and in parallel, without providing a unified execution environment or enforcing a consistent global transaction order.

Phala employs an event-sourced strategy, recording transaction inputs on-chain while executing contracts and maintaining state within off-chain TEEs \cite{yinPhalaNetworkSecure2022}. While suitable for simple, sequential inter-contract calls (e.g., Contract B executes only after Contract A completes), this approach lacks the flexibility required to support complex, result-dependent interactions (e.g., Contract A invokes B and relies on its output for further computation).

Therefore, existing solutions have not fully addressed private and efficient contract interoperability—a crucial capability for broader smart contract adoption, as exemplified by the insurance use case. We identify three fundamental design limitations that lead to this gap:

\begin{enumerate}
    \item Isolated Execution Environments: Confidential contracts are executed independently, each with separate keys and isolated off-chain nodes, lacking mechanisms for secure inter-contract communication and shared state access.

    \item Unreliable contract availability: In the event of off-chain node failure, contract redeployment typically requires manual node selection and full reinitialization, with execution resuming from inconsistent checkpoints across different contracts. Consequently, this leads to unpredictable service interruptions and significantly undermines the availability and stability of interdependent contracts.

    \item Transaction Order Inconsistency: Existing frameworks often process transactions from different contracts in parallel without enforcing a global execution order. This can lead to consistency issues during cross-contract interactions that involve interdependent state.
\end{enumerate}

To effectively support contract interoperability, a robust framework must address these three critical issues while maintaining—or enhancing—the privacy, efficiency, and liveness offered by prior approaches.

\section{Threat Model and Assumptions}
\label{sec:adversary}
Our goal is to enable any user to execute long-lived, interactive, and privacy-preserving smart contracts on public blockchains by leveraging off-chain TEE-enabled nodes for enhanced efficiency and confidentiality. To maintain a clear scope, we explicitly state our assumptions about relevant components and provide a detailed adversary model.

\subsection{Assumptions}
\subsubsection{Blockchain and On-chain Contracts}
RaceTEE is agnostic to the underlying blockchain platform and can operate atop any chain that satisfies a set of foundational security assumptions. We assume the blockchain behaves as a reliable, append-only ledger that guarantees the immutability, availability, and consistency of stored data through a robust consensus mechanism. Specifically, the platform must provide: (i) a globally consistent and totally ordered transaction history; and (ii) correct, deterministic, and verifiable execution of on-chain smart contracts according to its execution semantics (e.g., the Ethereum Virtual Machine). Additionally, we assume that RaceTEE’s on-chain components are correctly implemented and free from known vulnerabilities.

\subsubsection{TEEs}
RaceTEE is also TEE-agnostic and compatible with any TEE that provides reliable remote attestation, enforces execution integrity, and offers a dependable source of randomness. We trust the hardware manufacturer to provision TEEs correctly and ensure their foundational security properties.

While RaceTEE assumes that TEEs protect the confidentiality of their self-owned internal state, it does not require TEEs to be entirely immune to leakage of shared information among them—particularly in the presence of known sensitive information extraction attacks, including side-channel attacks \cite{vanschaikSoKSGXFailHow2024}. In contrast to prior work that assumes absolute confidentiality guarantees, we adopt a more practical threat model that acknowledges the existence of such vulnerabilities. Protection mechanisms against related attacks are considered orthogonal to our design but can be integrated as complementary safeguards.

We further assume that enclave-resident programs are correctly implemented and free from known exploitable bugs. Each TEE is expected to establish a secure and isolated execution environment for user-defined contracts, preventing unintended interactions or interference between contracts. However, our framework does not inherently protect against data leaks resulting from logical flaws in contract code, including intentional disclosure or faulty implementation.

\subsubsection{Cryptographic Specification}
RaceTEE relies on standard cryptographic primitives, including cryptographic hash functions, digital signatures, symmetric-key encryption, and public-key encryption algorithms. The framework is agnostic to the specific choice of algorithms, allowing flexibility based on deployment environments, regulatory requirements, or evolving cryptographic best practices. However, all selected algorithms must comply with well-established security standards and provide strong guarantees of confidentiality, integrity, authenticity.

\subsection{Threat Model}
We assume a strong adversarial environment in which all participants are untrusted and may deviate from the prescribed protocol to pursue their own interests. Adversaries are categorized into three distinct classes, which may arbitrarily collude.

\subsubsection{Malicious Users}
These adversaries may deploy arbitrary contracts or invoke functions maliciously to disrupt operations or extract confidential information. Representative attack vectors include deploying adversarial contracts, triggering denial-of-service (e.g., via excessive computation), or attempting unauthorized access to restricted functions.

\subsubsection{Malicious Off-chain Node Hosts}
These adversaries have full control over the host systems and the hardware running TEE instances. They may manipulate the communication interface between the TEE and its environment by forging, delaying, reordering, or suppressing inbound or outbound messages. They may also arbitrarily disconnect or terminate nodes. While they may in rare cases partially compromise confidentiality, especially concerning shared data, they are assumed incapable of violating the integrity of TEE execution or tampering with its internal state.

\subsubsection{External Adversaries}
These adversaries operate at the network layer and may intercept, delay, reorder, or modify protocol messages in transit, aiming to infer sensitive information or disrupt protocol execution. However, we assume they do not have full control over the network and that messages between honest participants are eventually delivered.

\section{Design}
\label{sec:design}
RaceTEE leverages TEE-enabled platforms as off-chain nodes to efficiently ensure smart contract confidentiality and execution integrity. To facilitate contract interoperability, RaceTEE eliminates direct interactions between users and off-chain nodes. Instead, it uses the blockchain as an intermediary to store checkpoints for automatic recovery at each block, maintain global transaction order for deterministic inter-contract execution, and provide a user-friendly, conventional interaction model. To further reduce on-chain storage costs, participating nodes form a peer-to-peer (P2P) network to store system and contract related data off-chain.

\subsection{Overview}
Fig.~\ref{fig:highLevelArchitecture} illustrates the high-level architecture of RaceTEE. As shown, RaceTEE consists of three main roles:

\begin{figure}[t]
  \centering
  \includegraphics[width=\linewidth]{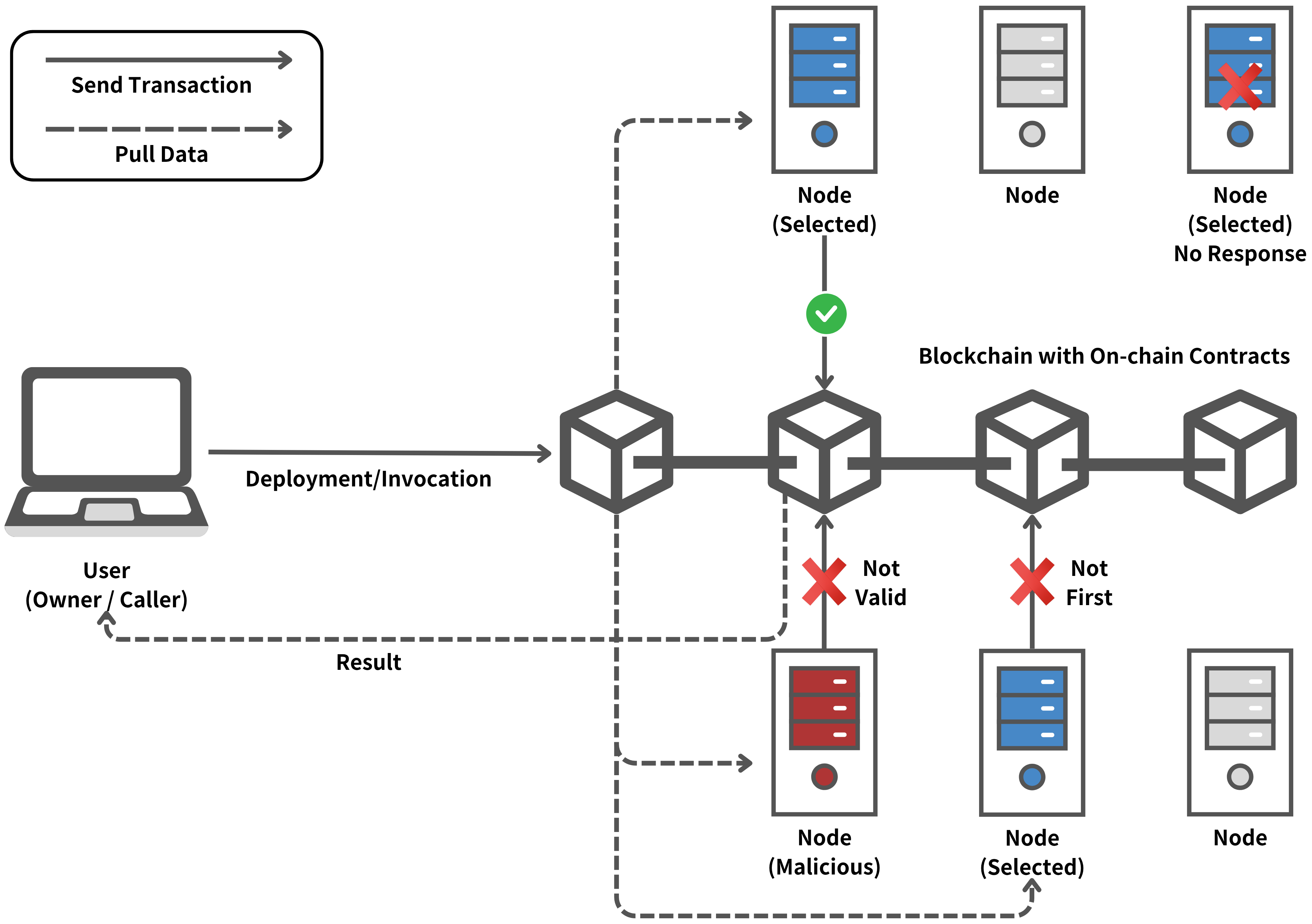}
  \caption{High-Level Architecture of RaceTEE}
  \label{fig:highLevelArchitecture}
\end{figure}

\begin{itemize}
    \item Blockchain with On-chain Contracts: RaceTEE employs two types of smart contracts on-chain: the \emph{management contract} ($MC$) and the \emph{program contract} ($PC$). The $MC$ is a single pre-deployed contract responsible for managing metadata, handling node registration and enforcing data integrity. In contrast, the $PC$ is deployed per confidential contract, enabling users to submit execution requests and retrieve results through standard interactions.

    \item Users: Users are classified as \emph{owners} and \emph{callers}. Owners deploy confidential contracts and manage execution privileges through an access control list (ACL), while callers invoke contract functions and retrieve results. Both roles interact with the system by submitting on-chain transactions. A user can hold both roles simultaneously.

    \item Off-chain Nodes: Execution is carried out by a decentralized network of TEE-enabled nodes forming a peer-to-peer (P2P) overlay network. Any entity with a TEE-enabled platform can join to provide computational resources. Additionally, these nodes contribute storage capacity, redundantly and collaboratively storing related data.
\end{itemize}
In summary, before contract deployment or execution, off-chain TEE-enabled nodes must first register with the RaceTEE on-chain contracts. To create or invoke a contract, users submit transactions containing encrypted code/data and payments to the blockchain. Once a transaction is included in a new block, all registered nodes monitor the transaction, and a randomly selected subset retrieves its contents, decrypts all transactions, and competitively executes the queued requests within their TEEs. After execution, the encrypted results are uploaded back on-chain. While multiple nodes may submit results, the on-chain contract verifies each result and accepts only the first valid submission, rewarding the corresponding node with the user’s payment. This mechanism encourages timely and correct computation.

\subsection{Design Challenges}
To support contract interoperability while ensuring correctness, liveness, and privacy, the following key challenges must be addressed.

\subsubsection{Unified Execution Environment}
To preserve contract confidentiality, it is common practice to run each contract in a separate off-chain node with distinct encryption keys, adopted by prior work. However, such isolation hinders inter-contract communication and complicates dependency management. Therefore, it is essential to build a unified execution environment that facilitates cross-contract communication while enforcing strict access control to prevent unauthorized access, thereby achieving both interoperability and privacy.

RaceTEE adopts a double-layer encryption scheme combined with an ACL. In addition to protecting each contract with its own keys, all off-chain nodes share a common key securely maintained within TEEs, which is used to further encrypt contract-specific keys and associated metadata. Before execution, every invocation undergoes a contract-specific ACL check to ensure that access permissions are properly enforced.

Specifically, during transaction execution, the TEE first checks the ACL to verify whether the invocation—originating from a user or another contract—is authorized. If permitted, the TEE uses the common key to decrypt the contract-specific keys and associated metadata, and then decrypts and executes the contract logic. This process iterates through the invocation chain until the entire call sequence completes.

However, sharing a common key among all TEEs raises concerns: since TEEs are not absolutely immune to confidentiality breaches, compromising this key could undermine the entire system’s privacy. Therefore, RaceTEE integrates a periodic key rotation mechanism that provides forward and backward secrecy,   limiting the impact of potential key leakage (see Section~\ref{subsubsec:keyRotation}).

Finally, RaceTEE complements this unified environment with competitive execution, described below, to eliminate execution isolation among off-chain nodes and ensure contract availability.

\subsubsection{Contract Consistent Availability}
Maintaining system liveness amidst off-chain node failures is a well-recognized challenge. Prior solutions typically fall into one of three categories:
\begin{itemize}
  \item Collateral-based: Off-chain nodes are required to lock a significant deposit, which is forfeited upon dropout. The penalty is then redistributed to compensate affected users \cite{dasFastKittenPracticalSmart2019b,renDeCloakEnableSecure2023}.
  \item Checkpoint-based: Off-chain nodes periodically upload encrypted snapshots of their internal state to the blockchain. Upon dropout, a replacement node resumes execution from the most recent checkpoint \cite{chengEkidenPlatformConfidentialityPreserving2019a,gaddamLucidiTEEScalablePolicyBased2023,liTrustingComputingService2023,duEPTEnhancingUser2023,yinPhalaNetworkSecure2022}.
  \item Replicated-state: Multiple off-chain nodes maintain replicated contract states. One acts as the primary executor, while the others serve as passive backups, ready to take over if the main node fails \cite{frassettoPOSEPracticalOffchain2023a,yuanShadowEthPrivateSmart2018}.
\end{itemize}

However, these approaches do not guarantee constant contract availability. In all cases, recovery requires reinitializing the contract environment and re-executing from a certain point to reconstruct the current state. This recovery time is variable and unpredictable, depending on contract complexity and customized recovery logic.

This uncertainty imposes a substantial burden on contract developers: if an upstream contract becomes unavailable during invocation, the downstream execution may stall or fail unpredictably. Consequently, developers are forced to implement complex exception handling logic and risk degrading user experience, especially in deep dependency chains where a single failure can propagate.

Inspired by Bitcoin’s proof-of-work model, we introduce a competitive execution model. All off-chain nodes concurrently attempt to execute transactions, but only the first node to produce a valid result is rewarded. As long as at least one honest node remains responsive, the system achieves effectively zero-latency recovery and maintains continuous availability of the contract.

Unlike traditional redundant on-chain execution, RaceTEE’s competitive approach does not require consensus from all nodes (e.g., it does not need to wait for a majority to complete computation and reach agreement through communication), thereby maintaining efficiency. Nevertheless, having all nodes execute every transaction introduces new challenges, such as wasted computational effort, centralization risk due to high-performance nodes, and increased user costs (as the rewards for successful execution must cover the cost of unsuccessful but performed computations). To mitigate these issues, RaceTEE employs a randomized node selection algorithm that restricts each execution round to a small, unbiased subset of nodes (see Section~\ref{subsubsec:nodeSelection}).

However, randomized selection can lead to rounds where no selected node responds, thereby reintroducing gaps in contract availability. To ensure progress and uphold the "anytrust" execution model (i.e., progress as long as at least one honest node responds), RaceTEE integrates a lightweight checkpoint mechanism. Each time a valid result is published on-chain, it is accompanied by a record indicating the most recently processed transactions. As a result, off-chain nodes can unambiguously determine which transactions have been completed and which remain pending. In the next round, nodes can resume from this consistent state, executing only unprocessed transactions.

This approach ensures that either all contracts are unavailable in a round (if no node responds) or all are available (if at least one valid result is returned), thus realizing what we term contract consistent availability.

In addition, to reduce on-chain checkpoint costs, RaceTEE publishes only a hash of each checkpoint on-chain as an anchor, while storing the full checkpoint data off-chain within the off-chain distributed storage (see Section~\ref{subsubsec:distributedStorage}).

\subsubsection{Global Transaction Order}
Prior work typically treats each transaction as an independent execution unit, resulting in unordered processing across contracts. This lack of coordination undermines cross-contract consistency and introduces race conditions, particularly when contracts share or depend on overlapping state variables. One might consider reordering transactions at off-chain nodes through peer communication. However, designing such a mechanism is challenging, error-prone, and incurs significant communication and computation overhead.

We argue that since the blockchain already employs sophisticated consensus mechanisms to enforce a deterministic transaction order within and across blocks, off-chain systems should leverage this existing property rather than reimplementing it. Accordingly, RaceTEE treats an entire block as a single atomic execution unit, thereby eliminating the need for additional off-chain transaction reordering logic. This approach significantly simplifies the system design while enhancing robustness and efficiency.

\subsubsection{Malicious Off-chain Node Hosts}
In TEE-based off-chain systems, malicious hosts pose a critical threat as they mediate communication between TEEs and the blockchain. A compromised host may fake, drop, reorder, or delay messages, undermining transaction correctness and freshness.

RaceTEE addresses these two issues separately. For correctness, TEEs validate each transaction’s authenticity by verifying its digital signature and the Merkle tree proof of inclusion (comparing the block header’s Merkle root with the transaction data), ensuring that only authentic blockchain data is processed inside TEE. This prevents a malicious host from forging a fictitious block containing fabricated transactions or altered content.

However, such verification cannot prevent the host from supplying an outdated (yet valid) block. Due to the absence of a trusted clock within TEEs and the asynchronous nature of block arrival, it is hard for TEEs to determine in real-time whether a received block is stale or not. To address this, RaceTEE adopts an optimistic post-verification strategy. 

Specifically, the TEE proceeds to process transactions under the assumption that the provided block is recent. Upon completion, the TEE attaches a freshness proof to its output. This proof binds the result to the exact block processed by the TEE: it includes the starting and ending block header hashes for that execution, both signed by the TEE. The result is then submitted to the on-chain $MC$, which verifies the proof against the current blockchain state. Only results derived from valid recent blocks are accepted.

As a result, if a malicious host feeds the TEE an outdated block, the resulting computation will be rejected by the on-chain verifier, and the attacker merely wastes its own resources without compromising system integrity.

\subsubsection{Compatibility and User-Friendliness}
Building a new ecosystem from scratch presents significant adoption challenges, therefore ensuring compatibility and a user-friendly experience is essential for practical deployment.

To avoid modifying the underlying blockchain consensus layer and to keep users unaware of off-chain complexity, RaceTEE introduces an $MC$ to handle system-level logic, and per-contract $PC$s to preserve the conventional user interaction model (see Section~\ref{subsubsec:onchainContracts}). This design eliminates the need to alter the underlying consensus and keeps off-chain complexity hidden from users. As a result, interacting with a RaceTEE confidential contract feels identical to interacting with a traditional contract.

\subsection{Workflow}
\begin{figure}[t]
  \centering
  \includegraphics[width=\linewidth]{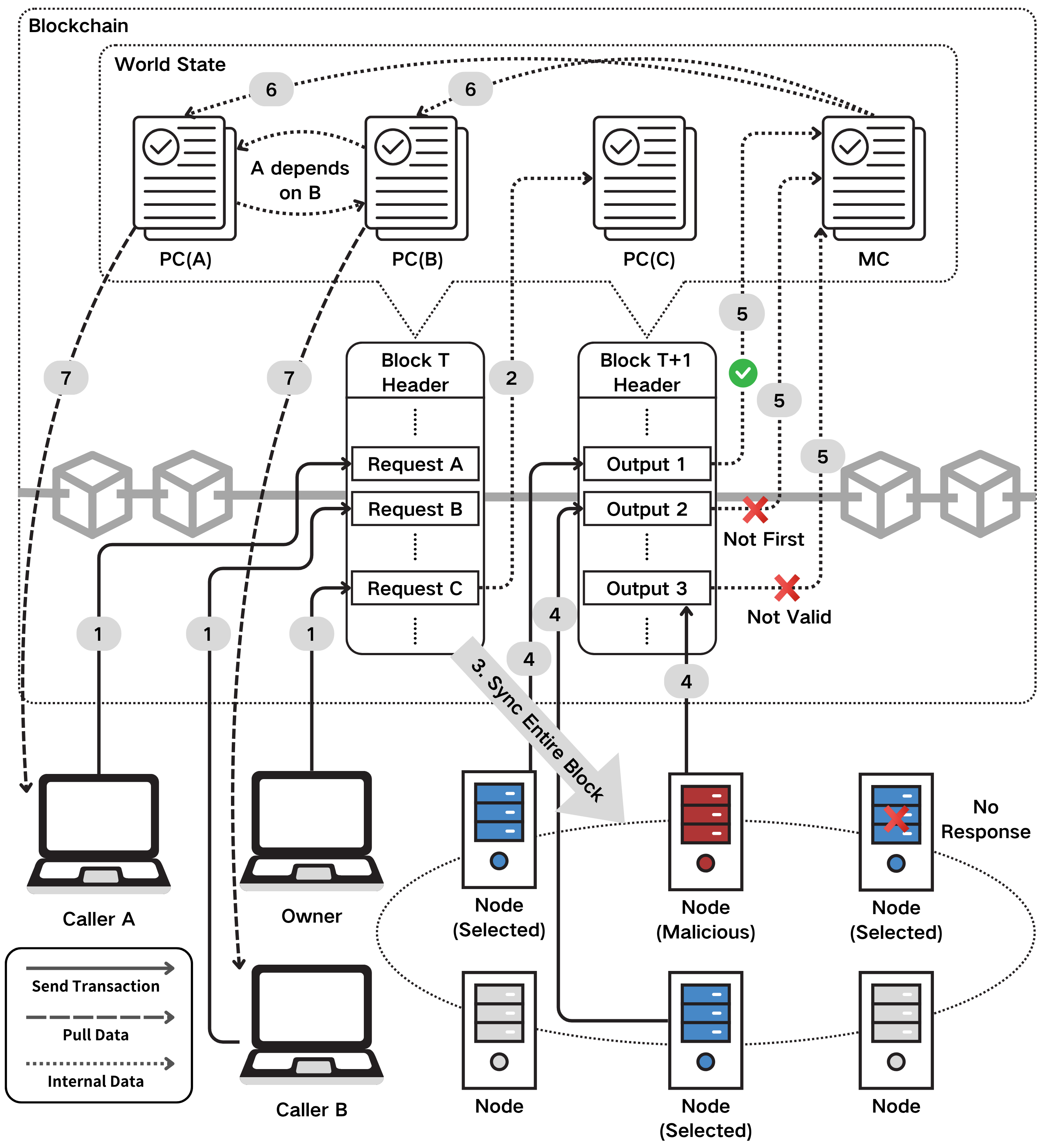}
  \caption{Execution Flow of RaceTEE}
  \label{fig:detailedArchitecture}
\end{figure}

Fig.~\ref{fig:detailedArchitecture} provides RaceTEE’s detailed workflow for a single round, with numbered steps corresponding to those in the figure. For brevity, we use $K$ to denote symmetric keys, $KP$ for asymmetric key pairs, $PriK$ and $PubK$ for private and public keys respectively, and $H$ for cryptographic hashes. We provide explanations for each notation both inline at its first occurrence and collectively in Table~\ref{table:notation} for ease of reference.
\begin{table}[b]
\caption{Explanation of Major Notations in RaceTEE}
\label{table:notation}
\centering
\scriptsize
\begin{tabularx}{\columnwidth}{lX}
\toprule
\textbf{Notation} & \textbf{Explanation} \\
\midrule
$MC$ & Management contract \\
$PC$ & Program contract \\
$LEB$ & Latest execution block number and hash \\
$Info_{\text{p}}$ & Per-contract management information \\
$KP_{\text{n}}$, $PubK_{\text{n}}$, $PriK_{\text{n}}$ & Asymmetric key pair for each TEE identity \\
$KP_{\text{tx}}$, $PubK_{\text{tx}}$, $PriK_{\text{tx}}$ & Asymmetric key pair for all requests encryption \\
$K_{\text{inf}}$ & Symmetric key for all contract management info \\
$K_{\text{code}}$ & Symmetric key per contract code \\
$K_{\text{st}}$ & Symmetric key per contract states \\
$K_{\text{res}}$ & Symmetric key per transaction result \\
$MKRP$ & Management key rotation period ($KP_{\text{tx}}$ \& $K_{\text{inf}}$) \\
$CKRP$ & Contract key rotation period ($K_{\text{st}}$) \\
$H_{\text{inf}}$, $H_{\text{code}}$, $H_{\text{st}}$ & Hash of contract management info, code, and states \\
\bottomrule
\end{tabularx}
\end{table}

\subsubsection{System Initialization}
Before operation, the on-chain $MC$ is deployed with initial system parameters, including the management key rotation period ($MKRP$) and the latest execution block reference ($LEB$). The first registered node generates the initial $K_{\text{inf}}$ (used to encrypt all contract-specific keys and metadata) and $KP_{\text{tx}}$ (used for encrypting user-submitted requests) within its TEE. It then publishes $PubK_{\text{tx}}$ and its TEE-derived identity $PubK_{\text{n}}$, along with the corresponding attestation quote on-chain for future reference and public verification.

\subsubsection{Off-chain Node Registration}
Off-chain TEE-enabled nodes must run a publicly auditable program, RacePro, which dynamically creates a sandboxed execution environment for user-defined confidential contracts. To join the system, an entity must load RacePro into its TEE, initialize it with the predefined parameters from $MC$, and execute it to generate $PubK_{\text{n}}$.

To establish trust, the newly initialized TEE undergoes remote attestation with a registered node, verifying its integrity and resulting in a signed endorsement of $PubK_{\text{n}}$. Upon successful attestation, the node registers with $MC$ by submitting the attestation signature alongside a deposit. Once registered, $PubK_{\text{n}}$ serves as the node’s identity, eliminating the need for repeated attestations in subsequent interactions.

\subsubsection{Contract Deployment and Invocation}
To deploy a contract, the owner first retrieves the latest $PubK_{\text{tx}}$ from the $MC$, then submits a deployment transaction to instantiate a $PC$ on-chain (Steps 1 and 2 from the owner). This transaction includes the encrypted contract code and configuration parameters (e.g., the ACL and the contract key rotation period, $CKRP$) as constructor inputs.

Once deployed, any authorized caller can submit an execution transaction to the $PC$, embedding two key components: (i) a newly generated symmetric key $K_{\text{res}}$, which the executing TEE will use to encrypt the result; (ii) call data encrypted under $PubK_{\text{tx}}$ (Step 1 from callers).

\subsubsection{Node Execution}
Upon block generation, off-chain nodes retrieve the block and verify its correctness within their TEEs (via RacePro) by validating the Merkle root in the block header against the included transaction data (Step 3). Each TEE then independently determines whether it has been selected for execution in the current round, based on the randomized selection algorithm (see Section~\ref{subsubsec:nodeSelection}).

Selected TEEs further check whether the $MKRP$ threshold has been reached. If so, the TEE generates a fresh $K_{\text{inf}}$ and $KP_{\text{tx}}$. While old keys remain valid for decrypting and processing transactions in the current round, all newly generated data is encrypted with the new keys.

Notably, the TEE treats the entire block as an atomic execution unit, processing all transactions in their globally defined order. For each transaction, it decrypts the input using $PriK_{\text{tx}}$ and executes it as follows:


\begin{itemize}
    \item Deployment Requests: The TEE generates $K_{\text{code}}$ and $K_{\text{st}}$ to protect the contract’s code and state, respectively. RaceTEE uses separate keys because the contract code remains static throughout its lifecycle, whereas the state evolves over time. This separation allows $K_{\text{st}}$ to be periodically rotated for protecting mutable contract state, while fixing $K_{\text{code}}$ for immutable code reduces the risk of ciphertext pattern leakage arising from repeated encryptions of identical content.

    The TEE then constructs $Info_{\text{p}}$, which includes the ACL, $CKRP$, an execution counter (initialized to 0), and the generated keys. This metadata is aggregated across all contracts and encrypted with $K_{\text{inf}}$, forming a unified management record.

    \item Invocation Requests: The TEE decrypts management information $Info_{\text{p}}$ using $K_{\text{inf}}$, and verifies the caller’s authorization against the ACL. If authorized, it retrieves the corresponding code and state hashes from the blockchain, then loads and decrypts the associated data from off-chain storage using $K_{\text{code}}$ and $K_{\text{st}}$. If the invocation involves inter-contract calls, the process recurses accordingly. Meanwhile, the execution counter is incremented with each invocation, and a new $K_{\text{st}}$ is generated once the $CKRP$ threshold is reached.

    Upon completion, the updated states and $Info_{\text{p}}$ are re-encrypted using $K_{\text{st}}$ and $K_{\text{inf}}$, respectively. The execution result is encrypted with the caller-specified $K_{\text{res}}$. If the request spans multiple contracts, individual records are generated for each interaction.
\end{itemize}

To mitigate risks arising from unbounded execution caused by either malicious denial-of-service (DoS) attempts or unintentional bugs, RacePro enforces a predefined execution time limit for each transaction. If this limit is exceeded, the execution is forcibly terminated, returning a “time-exceeded” result without updating the state or refunding any associated costs.

After processing all transactions in latest blocks, the TEE computes integrity hashes for the encrypted management info, contract code and states—denoted as $H_{\text{inf}}$, $H_{\text{code}}$, and $H_{\text{st}}$—based on the corresponding ciphertexts. Before submitting these hashes on-chain, the node disseminates the updated encrypted data to the other nodes via the P2P network. A randomly selected subset of nodes must then provide signed acknowledgments of receipt using their TEE-bound $PriK_{\text{n}}$. The TEE waits until the number of confirmations reaches a predefined threshold to ensure reliable off-chain data distribution (see Section~\ref{subsubsec:distributedStorage}). Once confirmed, the node signs the final transaction with its $PriK_{\text{n}}$ and publishes it on-chain, along with the block numbers and hashes indicating the processed range (Step 4).

\subsubsection{MC Updates}
These results (the execution outputs and state hashes) are included in the subsequent block by the miner. To ensure consistency and freshness despite potential variations in result arrival caused by network latency or malicious behavior, RaceTEE enforces strict validation within the $MC$, which accepts only the first valid execution output (Step 5).

Specifically, the $MC$ verifies each submission by checking the node’s TEE signature, and confirming that the reported execution range—defined by the start and end block numbers and hashes—matches the recorded $LEB$ and current blockchain status. Upon successful verification, the $MC$ updates the $LEB$, refreshes the stored $H_{\text{inf}}$, $H_{\text{code}}$, and $H_{\text{st}}$, and delivers the corresponding encrypted execution results to callers via the $PC$ (Step 6). While these results are publicly accessible on-chain, only the requesting caller holding corresponding $K_{\text{res}}$ can decrypt them, preserving privacy (Step 7). Finally, the $MC$ transfers the execution remuneration to the submitting node, incentivizing fast and correct computation.

\subsection{Building Blocks}
This section details the components of RaceTEE previously mentioned without in-depth discussion.

\subsubsection{On-chain Contracts}
\label{subsubsec:onchainContracts}

As shown in Algorithm~\ref{alg:mc}, $MC$ serves as the central repository for all off-chain node records and system-wide public metadata. It also functions as a trust anchor to ensure consistency of confidential contract states across participants.

\begin{algorithm}[b]
\small
\caption{$MC$ Pseudocode}
\label{alg:mc}
\SetKwProg{Fn}{Function}{:}{}
\KwData{
    $LEB$, $MKRP$, $PubK_{\text{tx}}$, \\
    $\mathit{NodeList} \gets$ Map(address $\to$ ($PubK_{\text{n}}$, $deposit$)), \\
    $\mathit{ProgList} \gets$ Map(address $\to$ $H_{\text{inf}}$), \\
    $\mathit{ProgCodes} \gets$ Map(address $\to$ $H_{\text{code}}$), \\
    $\mathit{ProgStates} \gets$ Map(address $\to$ $H_{\text{st}})$);
}
\Fn{Publish$(startBlock, endBlock, outputs, signature)$} {
    \If{$sender \in \mathit{NodeList}$ \textbf{and} $startBlock = LEB$ \textbf{and} $endBlock$ valid \textbf{and} $signature$ valid} {
        \ForEach{$o \in outputs$}{
            Parse $o$ as ($address$, $H_{\text{inf}}$, $H_{\text{code}}$, $H_{\text{st}}$, $K_{\text{res}}(result)$)\;
            Update $\mathit{ProgList}$, $\mathit{ProgCodes}$, $\mathit{ProgStates}$\;
            Record $(K_{\text{res}}(result))$ in corresponding PC\;
        }
        $LEB \gets endBlock$\;
        Send remuneration to sender\;
    }
}
\Fn{Register$(PubK_{\text{n}}, signature, deposit)$} {
    \If{$\mathit{sender} \notin \mathit{NodeList}$ \textbf{and} $signature$ valid}{
        $\mathit{NodeList}[\mathit{sender}] \gets (\mathit{PubK}_{\text{n}}, deposit)$\;
    }
}
\Fn{Withdraw$()$} {
    Delete $\mathit{NodeList}[\mathit{sender}]$ \textbf{and}
    Send back $\mathit{NodeList}[\mathit{sender}].deposit$\;
}
\end{algorithm}

It further provides three core functionalities for RaceTEE: Register, Withdraw, and Publish. The Register and Withdraw functions allow participants to join or leave the system with a refundable deposit, primarily to mitigate Sybil attacks. The Publish function is invoked when off-chain nodes submit execution results. It verifies the result’s authenticity by checking the TEE’s signature and ensures data freshness by validating the reported execution range. Specifically, including the start and end block numbers along with their hashes in the call data ensures that execution begins from the latest unprocessed transaction (the recorded $LEB$) and proceeds through the intended end block.

The $PC$ (Algorithm~\ref{alg:pc}) is deployed by a contract owner to represent a confidential contract on-chain, enabling users to interact with it through standard function calls. While any entity with the $PC$ address may submit invocation requests, execution privileges are strictly enforced within the TEEs, ensuring that only requests authorized by the ACL are processed.

\begin{algorithm}[t]
\small
\caption{$PC$ Pseudocode}
\label{alg:pc}
\SetKwProg{Fn}{Function}{:}{}
\Fn{Constructor$(PubK_{\text{tx}}(code), PubK_{\text{tx}}(config))$} {
    Record $(PubK_{\text{tx}}(code), PubK_{\text{tx}}(config), sender)$\;
}

\Fn{Execute$(PubK_{\text{tx}}(input), PubK_{\text{tx}}(K_{\text{res}}))$} {
    Record $(PubK_{\text{tx}}(input), PubK_{\text{tx}}(K_{\text{res}}), sender)$\;
}

\end{algorithm}

\subsubsection{Node Selection and Competition Mechanism}
\label{subsubsec:nodeSelection}
To balance execution overhead with liveness guarantees and scalability, RaceTEE adopts a randomized selection algorithm that restricts competition to a subset of off-chain nodes in each execution round. The algorithm is executed independently within each TEE without any inter-node communication, thereby minimizing coordination overhead. In addition, each node can determine its eligibility only after a new block is generated. This timing ensures unpredictability and fairness by preventing any entity from learning the selection outcome in advance.

To ensure fair selection, each TEE begins by determining the smallest increment (step) that is coprime with the total number of nodes $n$:
\begin{equation}
\text{step} = \delta + \min \left\{ \Delta \in \mathbb{N}_0 \,\middle|\, \gcd(\delta + \Delta,\ n) = 1 \right\}
\end{equation}

where \(\delta = \left\lfloor \tfrac{n}{c} \right\rfloor\), and \(c\) is the pre-defined number of nodes selected per round for execution. As node registration details, including count and order, are publicly available on-chain, each node can determine its order position \(o\) from its registration index in \(MC\) (i.e., if it was the i-th node to register, its position o = i). The selected node indices are then computed as:
\begin{multline}
o_k = \left(\bigl(S_r \bmod n) + k \cdot \text{step} \right) \bmod n, \\
\text{where } k \in \{0, 1, \dots, c{-}1\}
\end{multline}

It uses a per-round random seed $S_r$ to select $c$ distinct nodes out of $n$ in this round, preventing adversaries from dominating the system through consecutive registrations and ensuring fair distribution. For simplicity, the seed can be derived from the current block hash. To further mitigate potential miner manipulation, the randomness source can be replaced with a more secure on-chain source, such as a distributed randomness beacon \cite{sytaScalableBiasresistantDistributed2017,mengRondoScalableReconfigurationFriendly2025}, which is outside the scope of this paper.


The ratio $\tfrac{c}{n}$ plays a pivotal role in balancing computational overhead and response liveness per round. A higher value increases redundancy and competition, thereby reducing the expected failover latency to roughly one block interval. Conversely, a lower ratio means fewer nodes waste effort on the same task, reducing redundant computation. This parameter can be dynamically tuned based on observed request-to-response delays, enabling adaptive optimization of system responsiveness.

\subsubsection{Key Rotation}
\label{subsubsec:keyRotation}
To bound the impact of any leakage of shared information across TEEs, we rely on periodic key rotation to provide forward and backward secrecy, thereby limiting any exposure to a short time window.

To enhance security and limit the impact of key leakage, a key rotation mechanism is introduced to support forward and backward secrecy (i.e., confidentiality of past and future data even if a current key is compromised). This mechanism applies to three keys: $KP_{\text{tx}}$, $K_{\text{inf}}$, and $K_{\text{st}}$.

$KP_{\text{tx}}$ and $K_{\text{inf}}$ share a predefined main key rotation period ($MKRP$), determined before system initialization and specified by a block interval count. When a selected TEE detects that the on-chain $PubK_{\text{tx}}$ has not been rotated within the defined period, it generates fresh $KP_{\text{tx}}$ and $K_{\text{inf}}$ alongside transaction processing. These keys are securely broadcast to other peers through the P2P network, encrypted with each peer’s $PubK_{\text{n}}$, and require that a threshold number of peers acknowledge receipt before the system continues processing. Once the new $PubK_{\text{tx}}$ is fixed on-chain, all subsequent transactions and $Info_{\text{p}}$ must be encrypted using the new keys.

A smooth transition is ensured by allowing the old $PubK_{\text{tx}}$ to remain valid for a defined number of blocks, during which transactions encrypted under either key are accepted. Users specify the key version in the transaction input. After the transition window expires, older keys are retired, and transactions using them are rejected.

The contract key rotation period ($CKRP$) for $K_{\text{st}}$ is specified by the contract owner during deployment and set in the contract’s configuration based on the number of invocations. Each invocation increments a monotonic counter, and when it reaches the threshold, the selected node’s TEE generates a fresh $K_{\text{st}}$ to encrypt subsequent state updates. 

For traceability, the system can be extended to allow a contract owner (with proper identity verification) to periodically retrieve the current $K_{\text{st}}$ from a TEE and archive it securely outside the system. This enables the reconstruction of historical contract states when necessary.

\subsubsection{Off-chain Distributed Storage}
\label{subsubsec:distributedStorage}
RaceTEE employs a distributed storage scheme formed by the off-chain nodes’ P2P network to reduce on-chain storage costs. This storage layer replicates encrypted contract code, state, and $Info_{\text{p}}$, while committing only their hashes to the blockchain.

To ensure data availability, particularly against malicious nodes withholding data or benign nodes failing to receive or store them, all updated codes, states, and keys must be broadcast, and a threshold number of signed confirmations is required before execution results can be submitted on-chain. However, simply collecting a small number of confirmations introduces risks. For example, attackers controlling more than the threshold number of nodes can easily exploit the system. Conversely, increasing the number of required confirmations can degrade performance. To mitigate this, a scheme termed Random Subnet Threshold Signature (RSTS) is introduced below.

Suppose there are $n$ off-chain nodes in total. For any data broadcast intended for distributed storage, a TEE must collect at least $t$ confirmations before proceeding. To prevent attackers controlling $m$ nodes from signing confirmations solely within their controlled group, confirmations are drawn from a randomly selected subset of size $s$. Since network instability or unresponsive nodes may lead to lost confirmations, $s$ is set larger than $t$ to improve reliability. This ensures that confirmations are well-distributed across participants, reducing attack success probability $\epsilon$, given by:
\begin{equation}
\epsilon
=
\begin{cases}
0, & m < t, \\[6pt]
\displaystyle
\sum_{k=t}^{s}
\frac{\binom{m}{k}\,\binom{n - m}{s - k}}{\binom{n}{s}},
& m \ge t.
\end{cases}
\end{equation}

To ensure reliability, $\epsilon$ should be kept sufficiently low, where the success of an attack is deemed impractical. Our calculations show that this algorithm maintains low overhead even under challenging conditions. For example, suppose an adversary controls one-third of the nodes. Using a 90\% confirmation threshold ($t = 0.9 \times s$) in a network of 10,000 nodes, a randomly selected subnet of size $s = 38$ is sufficient to keep $\epsilon$ below $10^{-12}$. Additionally, a reputation system can be incorporated to filter out non-responsive nodes, allowing for a higher confirmation ratio while reducing the subnet size $s$. A detailed analysis of the interplay among these parameters is left for future work.

\section{Security Analysis}
\label{sec:security}
Based on the defined assumptions and threat model, we present a systematic analysis of RaceTEE’s correctness, liveness, and privacy.

\property{Correctness}{Once an execution result is fixed on-chain, the corresponding transaction’s effects (the state updates to all associated contracts) are guaranteed to be correct.}

All off-chain TEE-enabled nodes pre-load RacePro and undergo remote attestation to verify their identity with a previously registered peer. In this attestation, they obtain a signature on their randomly generated public key $PubK_{\text{n}}$. To register on-chain, this signature is verified by the management contract ($MC$) to ensure the node's legitimacy. Consequently, under the assumptions that (i) the TEE provides reliable randomness for $KP_{\text{n}}$ and keeps $PriK_{\text{n}}$ secure, and (ii) RacePro and the on-chain contracts execute correctly, no adversary can impersonate a valid registered node.

Once registered, nodes retrieve and process requests block by block. The TEE first verifies the block’s correctness by checking that the block header’s Merkle root matches the hash of the enclosed transactions. Requests are then executed in order, and the resulting outputs—including state updates, return values, block number, and block hash—are signed using the node’s private key $PriK_{\text{n}}$. This signature, along with a freshness proof, is subsequently verified on-chain by the $MC$ using the node’s $PubK_{\text{n}}$. Therefore, under the assumptions that the on-chain contracts and RacePro execute correctly, and the underlying blockchain remains secure, the above claim holds.

\property{Liveness}{Once a valid transaction is recorded on-chain, RaceTEE ensures its execution and result return within \( t \) rounds of block generation, assuming at least one benign off-chain node exists.}

Let $b$ be the block number where a valid transaction is recorded. In the first round, a subset of $c$ nodes is randomly selected to process it. If no node responds, another set of $c$ nodes is selected in the next round, ensuring that transactions from both block $b$ and block $b+1$ will be processed in that round. As time progresses, more nodes are selected, and all recorded requests up to the current block will eventually be processed, ensuring liveness.

In the worst case of $n$ total nodes with only one honest node, and $c$ nodes selected per round, the probability $\delta$ that a given request is processed within $t$ rounds is:
\begin{equation}
\delta = 1 - \left(\frac{\binom{n-1}{c}}{\binom{n}{c}}\right)^t = 1 -\left(1 - \frac{c}{n}\right)^t
\end{equation}

Since $c \leq n$, this probability converges to 1 as $t \to \infty$, proving that the system will eventually process every recorded request.

Moreover, economic incentives encourage greater participation from benign nodes while deterring non-responsive behavior from malicious ones. As the reward for computation in each round increases, more benign nodes are incentivized to participate, and rational but initially non-responsive malicious nodes may start responding to maximize rewards. 

Additionally, to deter denial-of-service attempts whereby a malicious user might submit an excessively long or even non-terminating computation, the system enforces an execution-time limit for each transaction. Any request exceeding this limit is automatically terminated, thereby ensuring that such adversarial attempts merely consume the attacker’s own resources and do not affect overall liveness.

\property{Privacy}{If a rational user deploys or invokes a confidential contract, its code, execution state, request inputs, and output results remain confidential throughout execution, assuming the TEE remains uncompromised.}

During deployment and invocation, contract code and execution inputs are encrypted with $PubK_{\text{tx}}$ prior to transmission, ensuring that only registered TEEs possessing the $PriK_{\text{tx}}$ can decrypt them. Upon execution, the contract code and state updates are further protected using $K_{\text{code}}$ and $K_{\text{st}}$, both of which are confined within the TEE. Similarly, the execution result is encrypted using $K_{\text{res}}$, a symmetric key shared exclusively between the TEE and the user. As a result, no plaintext information is exposed outside the TEE or to any entity other than the intended user. Under the assumption that the employed cryptographic primitives are secure, the design ensures end-to-end privacy.

Furthermore, recognizing information-leakage threats to TEEs, such as side-channel attacks, RaceTEE uses periodic key rotation that yields forward and backward secrecy, thereby restricting privacy leakage to a short time window, even in rare cases of key compromise.


\section{Implementation and Evaluation}
\label{sec:eval}
\begin{figure*}[t]
  \centering
  \includegraphics[width=0.8\linewidth]{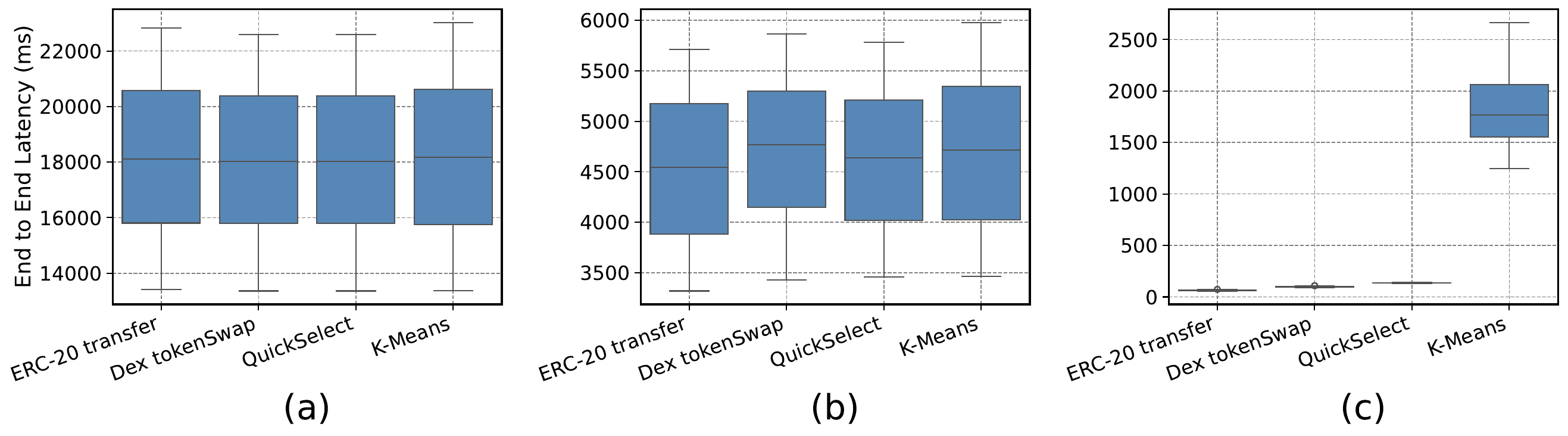}
  \caption{End-to-end latency of four contract functions across three execution environments: (a) a forked Ethereum blockchain with a 12-second block interval; (b) a high-performance blockchain with a 3-second block interval; and (c) a purely TEE-based RacePro execution without blockchain involvement.}
  \label{fig:latency}
\end{figure*}

\begin{figure*}[t]
  \centering
  \includegraphics[width=0.8\linewidth]{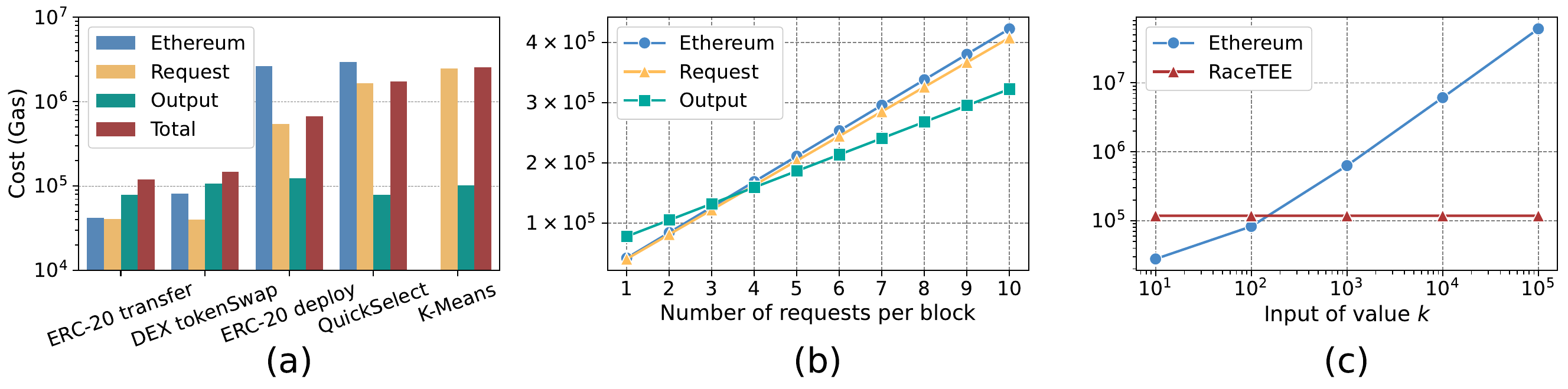}
  \caption{Cost comparison: (a) Cost of different requests, each executed once per block (log-scaled y-axis); (b) Cost of processing ERC-20 transfers as the number of requests per block increases; (c) Computation contract cost with varying input sizes (log-scaled x- and y-axes).}
  \label{fig:cost}
\end{figure*}
\subsection{Programming Model}
RaceTEE is designed to be TEE- and blockchain-agnostic. For evaluation, we implemented RaceTEE-SGXEth, using Intel SGX as the TEE and Ethereum as the blockchain due to their widespread adoption. To simplify SGX application development for RacePro, we leverage EGo, a framework that supports Golang running inside SGX \cite{ego}. Additionally, we ported the Ethereum Virtual Machine (EVM) into the TEE (using Go-Ethereum, Geth) to support on-chain contract execution. This setup also enables executing Golang programs within the enclave through the Yaegi interpreter. Furthermore, we developed an $MC$ and an abstract $PC$ in Solidity for system management and to facilitate the deployment of user-defined contracts. The implementation is available as open source\footnote{https://github.com/kerryzhangcode/RaceTEE}.

\subsection{Applications}
We develop five applications, ranging from basic Ethereum transactions to computationally intensive tasks, demonstrating the system's capabilities and versatility across various scenarios.

\subsubsection{Token Contract} One of the most widely used Ethereum contracts is the ERC-20 standard token contract. We implement a token contract based on this standard in Solidity, with the system automatically ensuring the confidentiality of token transfers and user balances.

\subsubsection{Decentralized Exchange (DEX)} Building upon our token contract, we implement a DEX, a fundamental component of decentralized finance (DeFi). Our DEX facilitates token swaps through liquidity pools, involving two token contracts and a primary exchange contract. This application not only demonstrates the system's ability to support complex inter-contract interactions in a practical scenario but also enhances privacy by concealing order details and liquidity pool information, mitigating front-running and sandwich attacks.

\subsubsection{Second-Price Auction} Using the ERC-20 token contract for payments, we implement a second-price auction where the highest bidder wins but pays the second-highest bid. Unlike the Ethereum mainnet, which struggles with maintaining bid confidentiality, our system preserves bidders’ privacy seamlessly.

\subsubsection{Quickselect Algorithm} To showcase RaceTEE’s efficiency with computationally intensive tasks, we implement the Quickselect algorithm to find the k-th smallest element in a dataset via a Solidity-based contract.

\subsubsection{K-Means Clustering for Machine Learning} Unlike previous Solidity-based applications, this one runs in Golang via the Yaegi interpreter. Machine learning algorithms are typically unsuitable for on-chain execution due to their computational complexity and general language constraints. However, our successful implementation of this widely used unsupervised learning algorithm demonstrates that complex workloads can be seamlessly ported and executed in RaceTEE, expanding potential use cases.

\subsection{Evaluation}
We deploy a test environment based on RaceTEE-SGXEth to evaluate its latency and cost. RacePro runs inside an Intel SGX enclave on a local machine running Ubuntu 24.04.2 LTS (Intel Core i7-8700, 16 GB RAM). Additionally, we run Ganache on the same machine to simulate the Ethereum blockchain. To model client interactions, we employ a MacBook Air 2022 (Apple M2, 16 GB RAM), simulating users sending contract deployment and invocation requests.

Our evaluation covers four types of function invocations: (i) a basic ERC-20 token transfer, (ii) a three-contract interaction for a DEX token swap, (iii) a computationally intensive QuickSelect operation to find the 1000th smallest number from a fixed set of 2048 randomly generated numbers, and (iv) a Go-based K-Means clustering algorithm, categorizing 1000 data points (5 dimensions) into 10 clusters over 100 iterations.

\subsubsection{Latency}
Latency is measured by sending each request once per block for 100 iterations, filtering out the top and bottom 10\% of results. As shown in Fig.~\ref{fig:latency}, overall latency is primarily determined by block generation time. Each request typically experiences a latency of one to two blocks from submission to result retrieval: the request is recorded on-chain in the first block, while the corresponding result is provided in the subsequent block. Given a reasonable execution time bound per request, we expect this latency to remain consistent in practice.

Notably, the system achieves lower end-to-end latency compared to orther designs that order transactions on-chain but execute them off-chain, where consensus ordering is crucial for enabling inter-contract interactions. Unlike existing approaches, RaceTEE does not require full transaction finalization prior to execution, which significantly reduces latency. In the event of a chain fork, all associated updates and results from discarded blocks are automatically invalidated.

In our current prototype, execution is relatively slow. We suspect this is due to certain implementation inefficiencies, including an unoptimized EVM port (from Geth), the overhead of the Go interpreter, and the need for multiple blockchain queries. With further engineering optimizations and refined execution paths, we anticipate significant performance improvements.

\subsubsection{Cost}
The cost of operating a high-performance VM with confidential computing on Azure is approximately \$1.61 per hour\footnote{Data as of March 4, 2025, for a machine with 32 vCPUs and 128 GiB RAM, and 1200 GiB storage in the East US region.}. Given a 12-second block generation time, this equates to approximately \$0.0054 per block, which is a negligible cost compared to on-chain transaction fees. For comparison, a simple ERC-20 transfer costs around \$0.44, with a gas consumption of 65,000 units, while a more complex DEX swap (Uniswap V3) costs around \$1.24, with 184,523 gas units consumed on the Ethereum mainnet\footnote{Data as of 15:42 GMT, March 4, 2025, from Etherscan.}. Thus, the primary costs come from the necessary blockchain interactions.

We measured the on-chain costs in gas units (to abstract away price fluctuations). Specifically, we analyzed the gas for one ERC-20 contract deployment and for each of the four types of function invocations described above. Since each request requires two transactions—one for the user’s request and one for the node’s output—our prototype introduces additional overhead for simple operations, as shown in Fig.~\ref{fig:cost}(a). For an ERC-20 transfer or a DEX token swap, our prototype incurs 2.8× and 1.8× the gas cost of direct Ethereum execution, respectively. However, for computationally intensive tasks, our prototype achieves significant cost savings, requiring only 58\% of the gas cost for a QuickSelect operation compared to on-chain execution.

Moreover, our prototype enables K-Means clustering, which is infeasible on-chain due to gas constraints, with estimated costs exceeding 10 million gas units on the Ethereum mainnet. This demonstrates RaceTEE’s potential for AI and complex computations that would otherwise be impractical on-chain.

To evaluate scalability, we analyze ERC-20 transfer costs when multiple requests are included in a single block. As shown in Fig.~\ref{fig:cost}(b), even with simple requests, the rate of increase in request and output costs remains lower than on the Ethereum mainnet, indicating that our prototype maintains reasonable economic overhead as user activity scales.

To further examine cost composition, we implemented a computation contract that takes an integer $k$ and performs a loop from 1 to $k$, summing or subtracting each number based on parity. Since input and output sizes remain constant, this isolates cost variations under increasing computational loads. As shown in Fig.~\ref{fig:cost}(c), Ethereum execution costs increase logarithmically, whereas our prototype maintains a constant gas cost regardless of $k$. This confirms that RaceTEE’s cost is primarily determined by input/output size rather than computational complexity, overcoming the challenge of supporting complex on-chain computation and thus broadening feasible applications.


\section{Conclusion}
\label{sec:conclusion}
In this paper, we identify and address the core challenges that limit off-chain confidential smart contract interoperability, a key requirement for enabling complex real-world applications. By enabling a unified execution environment, providing contract consistent availability, and supporting global transaction ordering, we introduce a new off-chain execution paradigm that is deterministically orchestrated according to the blockchain’s transaction order, without requiring additional consensus or synchronization among off-chain nodes. Based on this paradigm, we present RaceTEE, a novel off-chain framework for confidential smart contract execution that efficiently and comprehensively supports complex contract inter-contract interactions. Our evaluation demonstrates the practicality of RaceTEE across diverse scenarios and highlights its potential to advance decentralized applications.

As future work, we will investigate the interplay among system parameters to balance efficiency and security, conduct a formal security analysis, and present a detailed incentive model for competitive execution. Furthermore, the current prototype will be extended into a large-scale distributed system, enabling more extensive benchmarking and performance comparisons with existing platforms.

\bibliographystyle{IEEEtran}
\bibliography{references}

\end{document}